\documentclass[aps,prb,twocolumn,showpacs]{revtex4}
\usepackage{graphicx} 
\def\gapx{\lower 2pt \hbox{$\buildrel>\over{\scriptstyle{\sim}}$\ }}
\def\lapx{\lower 2pt \hbox{$\buildrel<\over{\scriptstyle{\sim}}$\ }}
\def\ph2{{\it p}-H$_2$}
\def\he4{$^4$He}
\def\beq{\begin{equation}}
\def\eeq{\end{equation}}
\def\Am3{\AA$^{-3}$}
\begin{document}

\widetext
\title{Superfluidity of $^4$He nanoclusters in confinement}
\author{Massimo Boninsegni} 
\affiliation{Department of Physics, University of Alberta, Edmonton, Alberta, Canada T6G 2G7}
\date{\today}

\begin{abstract}
Structure and superfluid response of nanoscale size \he4 clusters enclosed in  spherical cavities are studied by computer simulations.
The curved surface  causes the formation of well-defined concentric shells, thus imparting to the system a very different structure from that of free standing clusters. On a strongly attractive substrate, superfluidity is only observed at low density, in the single layer coating the inner surface of the cavity. If the substrate is very weak (e.g., Li), on the other hand,  a superfluid two-shell structure can form, whose physical properties interpolate between two and three dimensions. It is shown how experimental signatures of this physical behavior can be detected through measurements of the momentum distribution.
\end{abstract}
\maketitle
\section{Introduction}
Clusters  of \he4 comprising several tens of atoms (i.e., of characteristic size of a few nm) are perhaps the simplest quantum few-body system;  their investigation, both theoretical and experimental, has been pursued for a few decades now.
Early theoretical studies yielded considerable information on their structure,\cite{early,early2,chin,rk0}  Bose condensation,\cite{stringari,lewart}  excitations\cite{rk,ck} and their predicted superfluid behavior\cite{sindzingre89} at temperatures of the order of 1 K.
\\ \indent
Probing experimentally their physical properties  is obviously a difficult proposition, as pristine \he4 clusters  are very weakly bound, easily fragmented, e.g., on impact with a scatterer. Thus, although some  experiments were carried out, in which low energy \he4 atoms were scattered off \he4 clusters,\cite{toennies} most of what  is now known about their physics, chiefly their superfluid properties, has come from spectroscopic studies of  a single linear molecule (e.g., OCS)  embedded in them.\cite{grebenev,tang,xu}\\ \indent 
Despite the remarkable microscopic insight afforded by this approach, the presence of the embedded impurity, whose linear size is typically that of a few \he4 atoms, 
is expected to alter drastically the  structure of a cluster of such a small size. Specifically, a significant fraction of the \he4 atoms are predicted to bind to the foreign molecule, while the rest arrange in shells around it.\cite{mezzacapo} This is very different from the featureless structure of a pristine \he4 cluster.\cite{sindzingre89} The question therefore remains open as to whether one could probe experimentally the superfluid properties of a finite assembly of \he4 atoms in a way that does not make use of a foreign impurity, and preserves at least the full  rotational invariance of the cluster.
\\ \indent
Aside from possible experiments on free standing clusters, with the aforementioned difficulties that they entail, one possibility that does not seem to have been given much consideration so far is that of studying \he4 clusters enclosed in nanometer size cavities, e.g., those of a suitably chosen porous material. For example, one could think of adsorbing \he4 inside zeolites, whose pore diameter is typically of the order of a nm,  and therefore may accommodate clusters of a few atoms. \cite{note} One could therefore think of investigating the physics of the clusters, e.g., by performing neutron scattering experiments on the system, on the assumption that the bulk of the signal should come from helium confined inside the (relatively regularly shaped) cages. Obviously, the presence of a confining surface is likely to affect significantly the structure and physical behavior of the cluster, compared to those of a free one.  For example, one may expect superfluidity to be suppressed, with crystallization originating at the surface of the cavity, and then extending to adsorbed layers.
\\ \indent
While a classical hard-sphere fluid in spherical confinement has been the subject of much theoretical investigation,\cite{carignan,woods,chui91,gonzalez} surprisingly little is known about the effect of this type of confinement on the superfluid transition of \he4. A number of studies of superfluidity of \he4 in narrow cylindrical channels have been carried out, \cite{boninsegni07,delma,glyde,po} typically aimed at modeling  \he4 in porous media, or in the confines of carbon nanotubes.\cite{crespi,amj}
Theoretical studies of adsorption of \he4 on spherical substrates  have mostly focused on layering on the outer surface of macromolecules such as fullerenes;\cite{hernandez,urrutia,park14,hernandez14,park15} however,  the superfluid properties of \he4 clusters confined inside spherical cavities of nanoscale size remain largely unexplored. A recent theoretical study of parahydrogen clusters confined inside spherical cavities of nanoscale size has yielded surprising evidence of enhanced superfluid response, with respect to free clusters,\cite{tokunbo} showing that the curved surface can affect in nontrivial ways the behaviour of   the system.
\\ \indent
In this work, we carried out a theoretical study of structural and superfluid properties of clusters of \he4 confined inside a spherical cavity of diameter 2 nm. Specifically, we performed equilibrium Quantum Monte Carlo simulation of a reasonably realistic model of the system at low temperature (down to $T=0.5$ K).
We considered the two limiting cases or strong and  weak substrate, in order to differentiate geometrical effects from those arising from the specific nature of the adsorbing medium.
Our purpose is twofold: first, we aim at gaining some further insight in the effect of nanoscale confinement on the physical properties of \he4 clusters; second, we wish to furnish some definite theoretical predictions,  helping in the design and interpretation of experiments on \he4 adsorbed inside porous media whose open volume consists primarily of (quasi)spherical cavities of characteristic size of few nm.
\\ \indent
As expected, confinement drastically alters the structure of the clusters, in comparison to free standing ones, imparting clusters a well-defined shell-like structure; nevertheless, superfluidity is resilient, and still observable within such tight confines at temperatures of order 1 K, under relatively broad conditions. The character of the superfluid response changes considerably, depending on the strength of the substrate.
On a strong substrate, a quasi-2D superfluid film coats the surface, undergoes crystallization as the density increases, and no reentrant superfluid phase is seen as a second layer (i.e., a concentric shell) forms. On a sufficiently weak substrate, on the other hand, the system forms two concentric superfluid shells, reminiscent of what seen for parahydrogen in Ref. \onlinecite{tokunbo}. Signatures of such different physics can be detected in the single-particle momentum distribution, probed by neutron scattering, a technique which has offered a great deal of insight into the physics of \he4 and other quantum fluids (like parahydrogen) in the confines of porous media.\cite{azuah,albergamo0,albergamo,bossy}
\\ \indent
In the next section, we introduce the mathematical model of the system, and provide some basic details of the simulational technique utilized. We then illustrate our results and outline our conclusion in the following two sections.
\section{Model and Methodology}
Our system of interest
is an ensemble of $N$ \he4 atoms, regarded as point particles of spin zero, enclosed in a spherical cavity.
The quantum-mechanical many-body Hamiltonian is the following:
\begin{equation}\label{one}
\hat H = -\lambda\sum_{i=1}^N \nabla_i^2 + \sum_{i<j} V(r_{ij}) + \sum_i U({r}_i)
\end{equation}
Here, $\lambda=\hbar^2/2m$, $m$ being the mass of a \he4 atom,  ${\bf r}_i$ is the positions of the $i$th atom, measured with respect 
to the center of the cavity (set as the origin), $r_{ij}\equiv |{\bf r}_i-{\bf r}_j|$,  $V$ describes the interaction of a pair of atoms, whereas $U$ describes the confinement of each atom inside the cavity. For $V$ we used the accepted Aziz potential,\cite {aziz} whereas for $U$ we used the expression\cite{gatica}
\begin{equation}\label{potl}
V(r)=2D \biggl \{\frac{b^9 F(x)}{(1-x^2)^9}-\frac{6b^3}{(1-x^2)^3}\biggr\},
\end{equation}
where $x\equiv r/R$, $F(x)=5+45x^2+63x^4+15x^6$, $b\equiv(a/R)$ and $a$ and $D$ are two parameters that are adjusted to reproduce the main \he4 adsorption features of specific substrates. \\ \indent 
Eq. (\ref{potl}) is merely the extension to the case of a spherical cavity of the so-called ``3-9" potential, describing the interaction of a  particle with an infinite, planar substrate.
$D$ is  the depth of the attractive well of the potential experienced by a \he4 atom in the vicinity of the substrate, whereas $a$ is essentially the distance of closest approach of a \he4 atom to the substrate.
\\ \indent
As a model of \he4 in the confines of, e.g., the cages of zeolites, Eq. (\ref{one}) clearly contains simplifications, chiefly the fact that the cavity is regarded as perfectly spherical and smooth. However, it allows us to address the physical question that we wish to pose here, namely the effect of confinement on the superfluid response of a \he4 cluster. Equivalent, or even simpler models (e.g., cavities with hard walls) have been utilized to study structure of  $^4$He and classical fluids in confinement.\cite{delma,chui91}
\\ \indent
We obtained results for a cavity of radius $R = 10$ \AA, but with two distinct choices of the parameters $D$, $a$ in Eq. (\ref{potl}), corresponding to very different adsorption properties. The first choice, henceforth labeled with Li, has $D$=17.87 K and $a$=3.76 \AA; these values are apt to describe the interaction of a \he4 atom with a Li substrate, one of the most weakly attractive known, on which \he4 is predicted to form a superfluid monolayer at low temperature.\cite{io}
\\
The second choice, namely $D$=100 K and $a$=2.05 \AA, is roughly in the ballpark of what one would expect for \he4 atoms near a silica substrate;\cite{treiner} thus, we shall henceforth refer to the scenario described 
by this parameter set as 
glass. The considerably deeper well, together with a much shorter range of the repulsive core, render the ``glass" much more attractive to \he4 atoms than the Li cavity. Obviously, in neither case do we aim at reproducing accurately 
any realistic interaction (which would require more elaborate 
functional forms anyway).
Rather, our aim is that of investigating opposite ends of the adsorption continuum.  
\\ \indent
We studied the low temperature physical properties of the system described by Eqs. (\ref {one}) and (\ref{potl})  by means of first principle computer simulations based on the Worm
Algorithm in the continuous-space path integral representation.\cite{worm,worm2} Specifically, we used a variant of the method which allows one to perform calculations in the canonical ensemble (i.e., fixed $N$).\cite{mez,mez1} Because this well-established computational methodology is thoroughly described elsewhere, we shall not review it here. The most important aspects to be emphasized here, are that it enables one to compute thermodynamic properties of Bose systems at finite temperature, directly from the microscopic Hamiltonian, in particular  energetic, structural and superfluid properties of the confined \he4  fluid, in practice with no approximation. Technical details of the simulation are standard, and we refer the interested reader to Ref. \onlinecite{worm2}. All of the results reported here were obtained with a value of the imaginary time step 
$\tau=1/640$ K$^{-1}$, with a high-temperature approximation for the many-particle propagator accurate up to order $\tau^4$.

\section{Results}
\subsection{Energetics}
\begin{figure}  [t]         
\centerline{\includegraphics[scale=0.34]{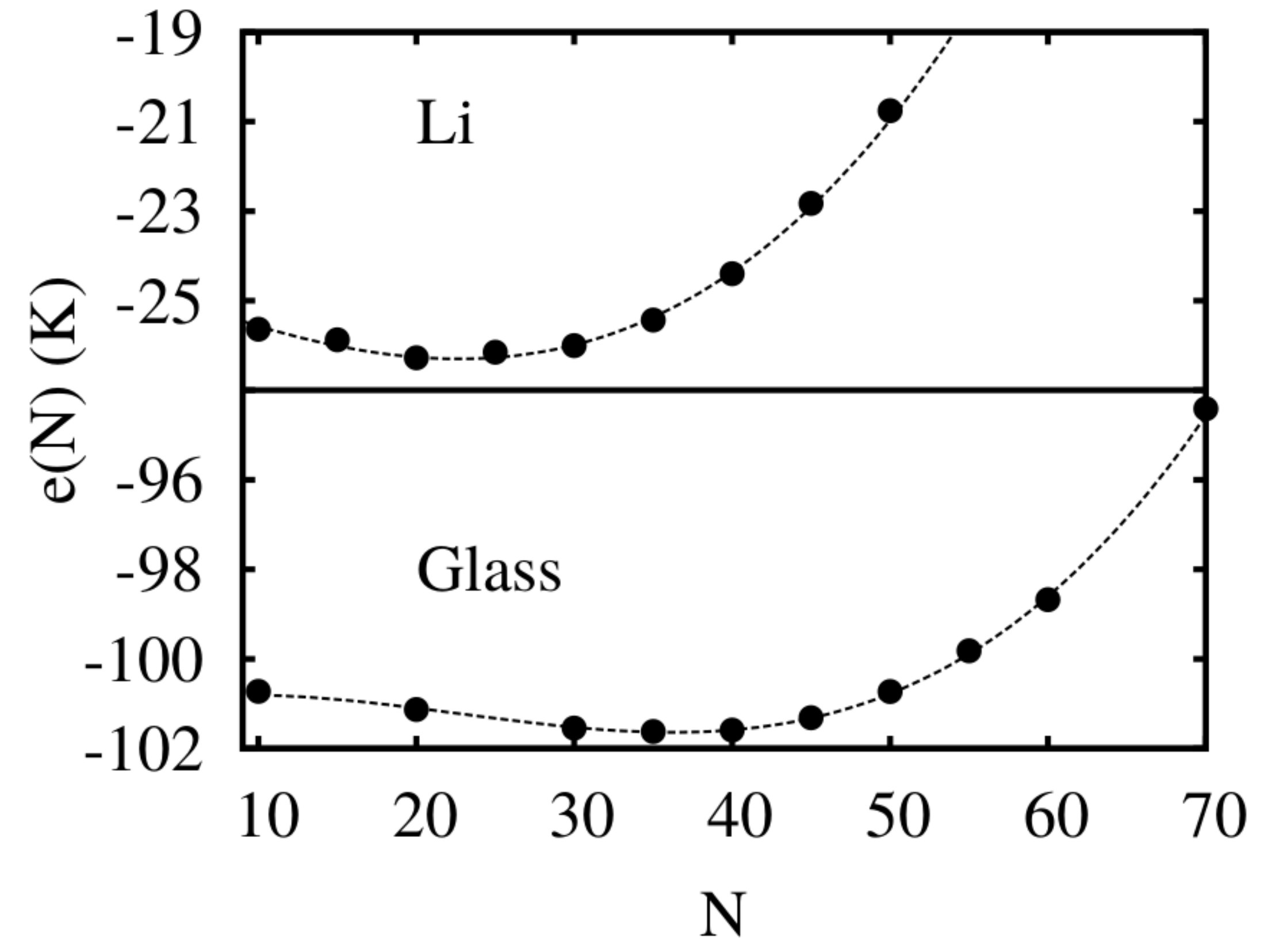}} 
\caption{Energy per \he4 atom $e$ (in K) versus number $N$ in the $T\to 0$ limit, inside
a Li (upper panel) and glass (lower panel) cavities of radius 10 \AA. Dashed lines are polynomial fits to the data.
Statistical errors are at the most equal to symbol size.}
\label{ene}
\end{figure}

We begin the illustration of the results of the simulations by discussing the computed energetics. We obtain ground state estimates by extrapolating to $T=0$ low temperature results. In pratice, we find that
energy values, as well as radial density profiles, remain unchanged, on the scales of the figures shown here, below 
$T \lesssim$ 1 K. 
\\ \indent
Fig. \ref{ene} displays the energy per \he4 atom (in K) as a function of the number of atoms in the cavity. Both curves feature minima at specific numbers of atoms, which correspond to the minimum filling of the cavity.  Polynomial fits to the data shown in Fig. \ref{ene} (which, for a glass cavity include energy estimates up to $N$=110, not shown for clarity in the figure), yield
a minimum for a Li (glass) cavity at approximately $N=22$ ($N= 38$)  \he4 atoms, at an energy close to $-26.3$ K ($-101.5$ K), which can be compared to the value $-9.6$ K ($-65$ K)  yielded by the corresponding 3-9 potential for a flat substrate.\cite{io,ancoraio}
Stronger atomic binding  arises from the curved confinement, and it is interesting to note that the enhancement is proportionally 
greater for the weaker substrate. \\ The same fits show that the chemical potential, obtained as $\mu(N)=e(N)+N (de(N)/dN)$ becomes comparable to that of bulk superfluid \he4 ($\sim -7.2$ K) for $N\approx 45$ for a Li cavity, and $N\approx 85$ for glass. This estimate is relevant to possible experiments in which a porous matrix in which \he4 is adsorbed is held in thermal contact with a bath of superfluid \he4. Inside a glass substrate, at those physical conditions, we estimate the kinetic energy per \he4 atom to be $38.0\pm0.5$ K. This value is in the same ballpark as the most recent experimental estimates\cite{andreani} of  the mean atomic kinetic energy of \he4 adsorbed in cylindrical nanopores of diameter 24 \AA, but the very large ($\sim$ 20\%) uncertainty quoted therein renders such a comparison scarcely meaningful. In general, one might expect that the kinetic energy to be mostly affected by the characteristic size of the confining medium, not so much by its specific geometry. For a cluster of 45 atoms inside a Li cavity, we find the kinetic energy per \he4 atom to be $21.33\pm 0.03$ K at $T=1$ K.
\begin{figure}  [h]        
\centerline{\includegraphics[scale=0.36]{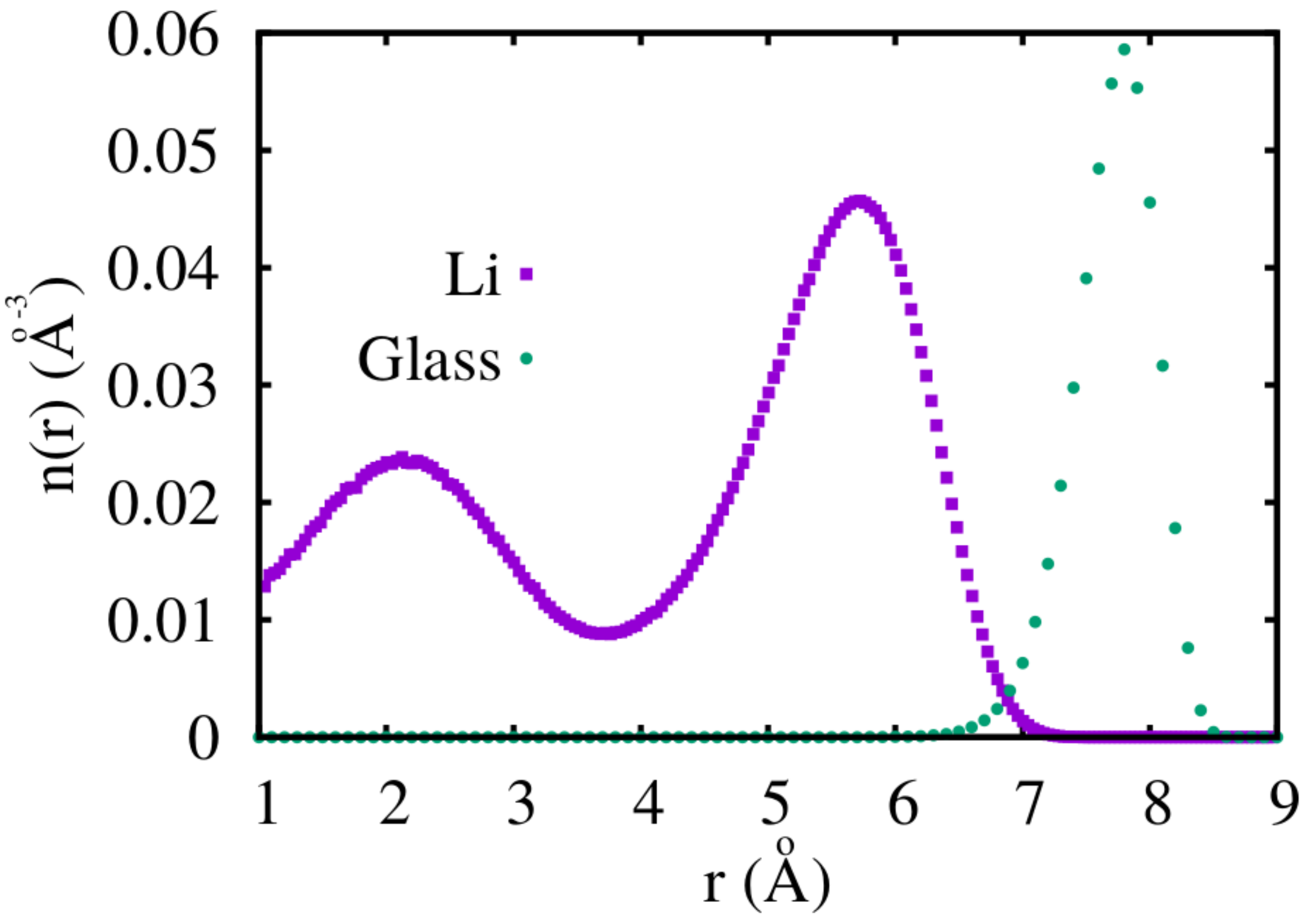}} 
\caption{Radial density profiles at $T$=1 K for a cluster of 25 \he4 atoms inside a Li cavity (boxes) and for 35 \he4 atoms inside
a glass cavity (circles). The origin is at the center of the cavity. Statistical errors are at the most equal to symbol size.}
\label{rhoeq}
\end{figure} 
\begin{figure}  [h]        
\centerline{\includegraphics[scale=0.36]{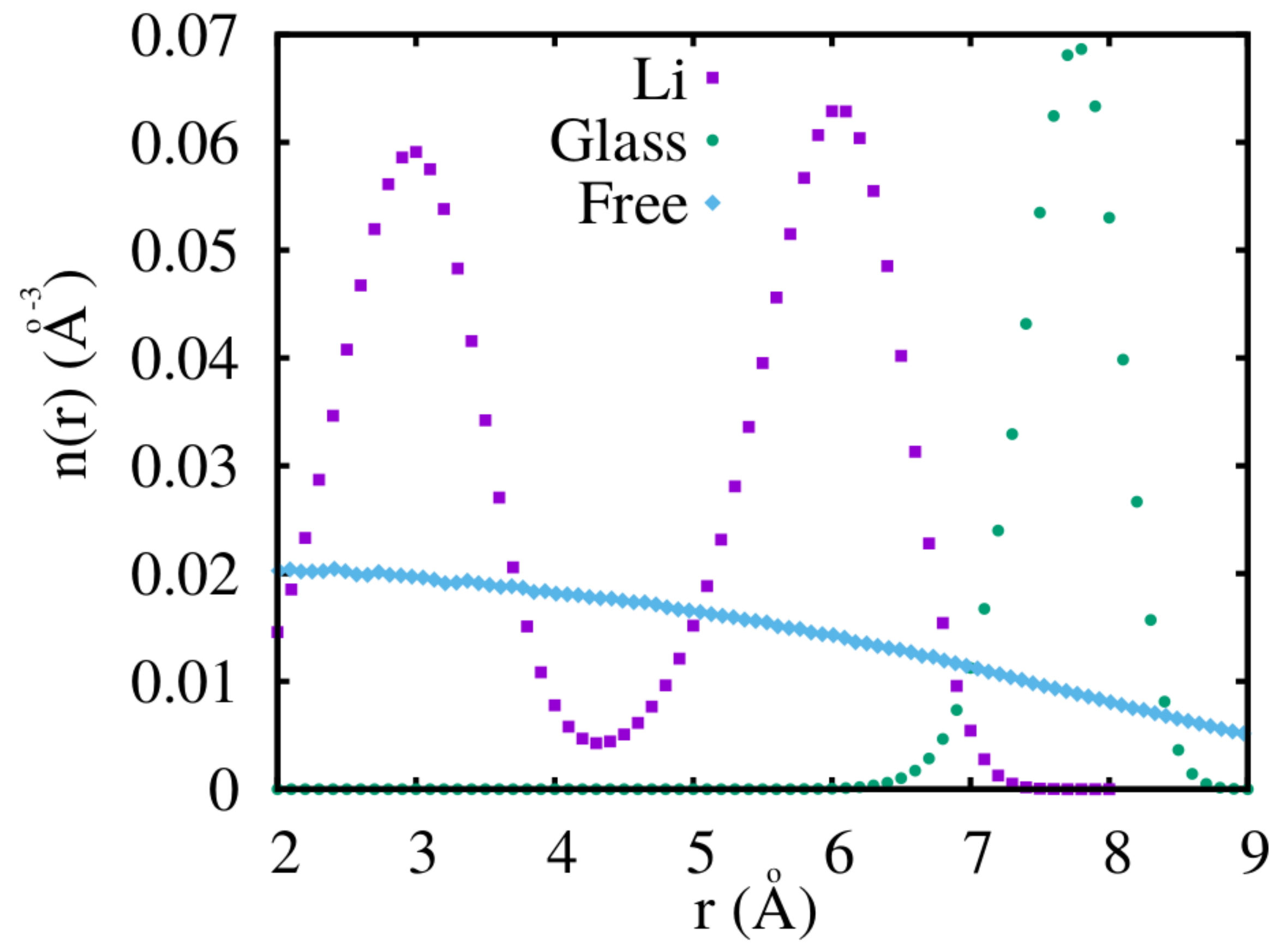}} 
\caption{Radial density profiles at $T$=1 K for a cluster of 45 \he4 atoms inside a Li cavity (boxes) and
a glass cavity (circles). The origin is at the center of the cavity. Also shown is the result for a free \he4 cluster (diamonds), computed with respect to its center of mass. Statistical errors are at the most equal to symbol size.}
\label{hig}
\end{figure} 
\subsection{Structure}
Fig. \ref{rhoeq} shows radial density profiles $n(r)$, computed with respect to the center of the cavity,  for a cluster of 25 atoms inside the Li cavity (boxes), and for one of 35 inside the glass one (circles), i.e., in both cases close to the equilibrium filling. Clearly, the arrangement of the particles in the two cases is very different, reflecting the different strengths of the adsorption potentials. Inside the more attractive glass cavity, \he4 forms a single thin film (a spherical 
shell) coating the surface, whose effective 2D density at the minimum ($N=38$) can be estimated to be $\sim 0.05$ \AA$^{-2}$, based on the position of the peak of $n(r)$.  This value is only slightly greater than the \he4 2D equilibrium density.\cite{pollock} The structure of the adsorbed film comprises a single shell. and this remains the case for $N\lesssim 80$, at which point the second shell begins to appear. This corresponds to a surface coverage of approximately 0.100 \AA$^{-2}$ where second layer promotion begins to occur, quantitatively consistently with what observed on a flat substrate.\cite{ancoraio}
\\ \indent
Inside a Li cavity, on the other hand, two largely overlapping, floppy concentric shells form, as atoms experience a repulsive core extending over 3 \AA\ away from the surface. Thus, even though \he4 does form a monolayer on a flat Li substrate,\cite{io} inside a nm-radius spherical cavity the energetically favored arrangement is droplet-like. This result mimics what observed for parahydrogen clusters in the same setup, on a weakly attractive substrate.\cite{tokunbo} 
Fig. \ref{hig} shows density profiles for systems comprising $N=45$ \he4 atoms inside the two cavities, at $T=1$ K. The change is minimal on a glass substrate, as the \he4 film coating the surface of the cavity becomes compressed, whereas the two shells inside the Li cavity become noticeably more rigid, especially the inner one, and a clearer demarcation between them emerges, although substantial  atomic overlap in the region between the two shells remains. The structure of these clusters is clearly very different from that of a free cluster, as illustrated in Fig. \ref{hig}. A free \he4 cluster is very losely bound, featureless and extends out all the way to approximately 12 \AA\ from its center of mass.\cite{unp} The absence of structure for the free cluster is in agreement with previous finite temperature simulations.\cite{sindzingre89}
The interesting issue, of course, is how confinement, which modifies so significantly their shape, affects the superfluid properties of these clusters, and whether one might be able to probe them experimentally.
 \begin{figure}  [t]        
\centerline{\includegraphics[scale=0.36]{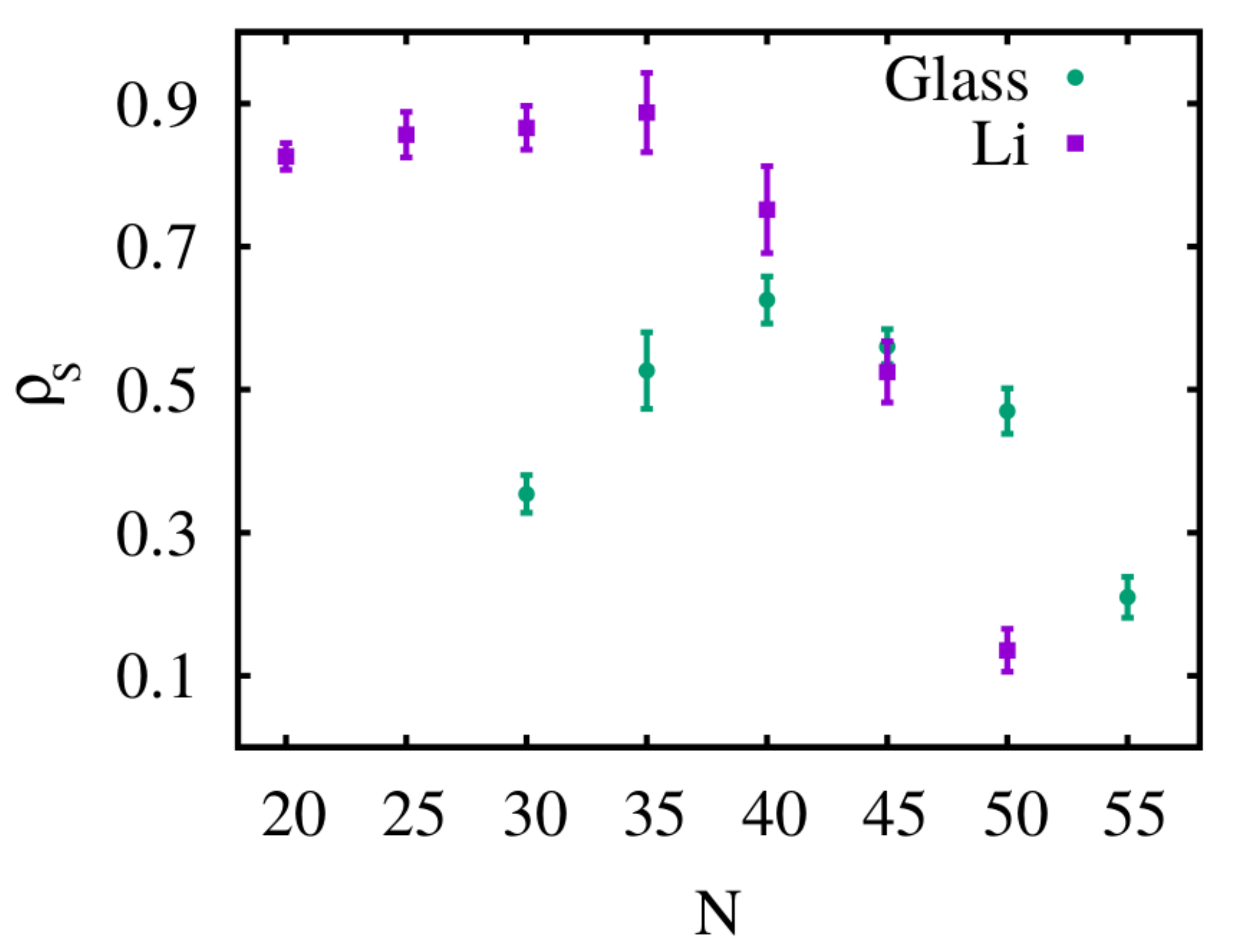}} 
\caption{Superfluid fraction at $T$=1 K of clusters of varying numbers of \he4 atoms enclosed inside a Li (boxes) and glass (circles) cavity of radius $R=10$ \AA. When not explicitly shown, statistical errors are at the most equal to symbol size.}
\label{rhos}
\end{figure} 
\subsection{Superfluidity}
Fig. \ref{rhos} shows the  superfluid fraction $\rho_S$ computed using the well-known``area" estimator\cite{sindzingre89} at temperature $T=1$ K for clusters of varying number of atoms, starting from $N$=20, at the lower limit of thermodynamic stability for a Li cavity, to $N=55$.  We also computed the same quantity for free \he4  clusters comprising $N=45$ and 55 atoms, both yielding approximately 82\%, i.e., essentially the same value reported by Sindzingre {it et al.} for $N$=64 at this temperature.
\\ \indent
The superfluid response is quantitatively different inside the two cavities. Specifically, on the weakly adsorbing Li substrate the two-shell structure that forms (Figs. \ref{rhoeq},\ref{hig})  displays a superfluid fraction in excess of 80\%, remarkably constant (within statistical uncertainties)  for $20\le N \lesssim 40$, declining rapidly above $N=40$ and becoming negligible for $N \gtrsim 50$, when the cluster  acquires a solid-like structure with the appearance of atoms at the center of the cavity. Inside a glass cavity,  the single film coating the surface has a finite superfluid response at the equilibrium coverage, quantitatively similar to that inside a Li cavity, for the same number of atoms. As the number of atoms is increased, the film is compressed and the superfluid response weakens, becoming negligible for $N\gtrsim 55$, which corresponds to a 2D coverage around 0.072 \AA$^{-2}$, in quantitative agreement with the location of the $T=0$ melting density in 2D.\cite{gordillo} Indeed, even inside a glass cavity of such narrow diameter the physics of the adsorbed \he4 is largely 2D. 
 \begin{figure}  [h]        
\centerline{\includegraphics[scale=0.34]{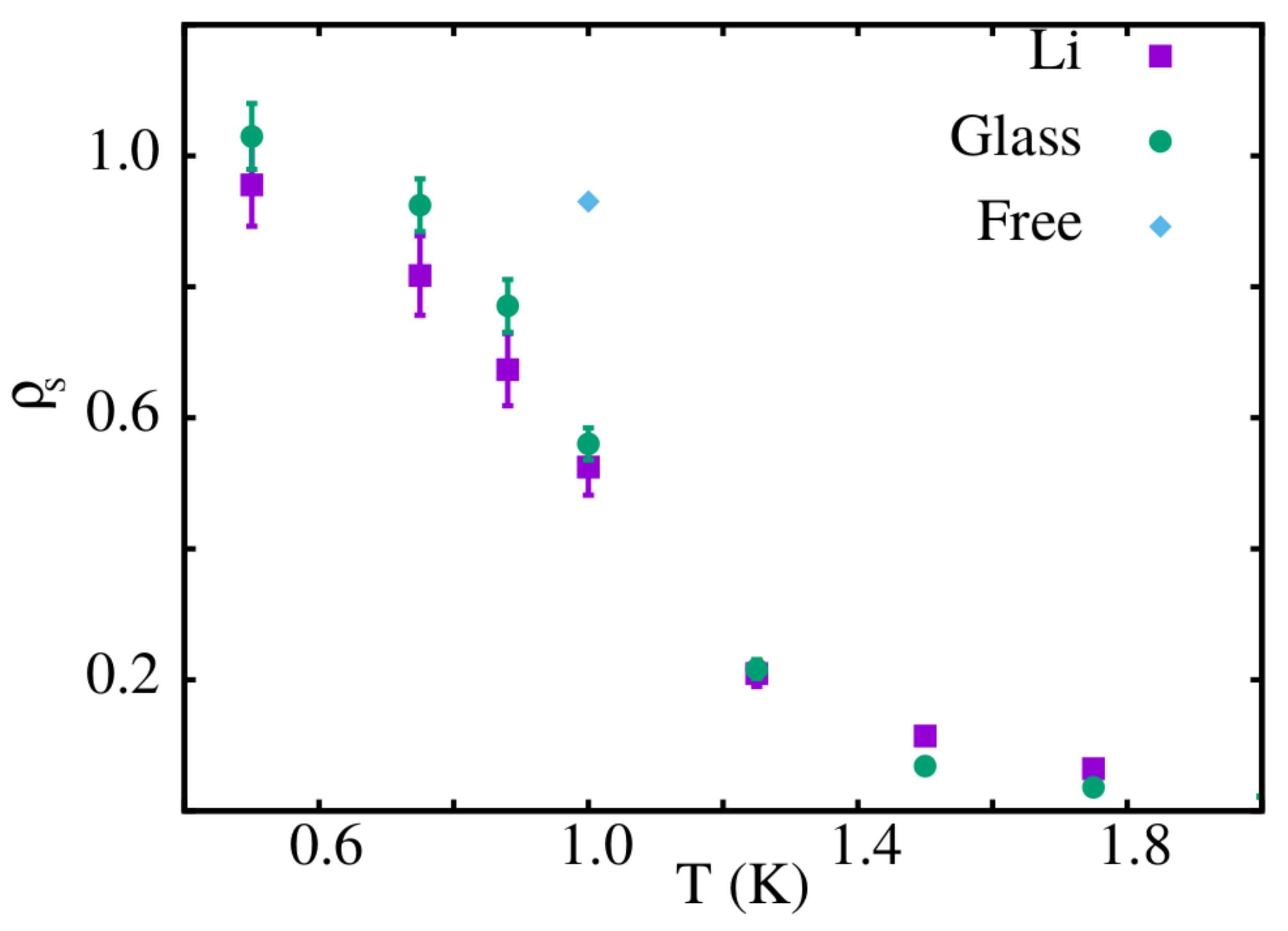}} 
\caption{Superfluid fraction $\rho_S(T)$ versus temperature for a cluster of $N=45$ \he4 atoms enclosed inside a Li (boxes) and glass (circles) cavity of radius $R=10$ \AA.  Also shown is a result for a free \he4 cluster (diamond), computed with respect to its center of mass. When not explicitly shown, statistical errors are at the most equal to symbol size.}
\label{rsofT}
\end{figure} 
\\ \indent
No reentrant superfluidity at higher density is seen  in either cavity; on the glass substrate, that means that inner shells are not superfluid, at least not at the lowest temperature considered here, namely $T$=0.5 K. This suggests that a glassy confining medium of the characteristic size considered here may be too tight for superfluidity to be observed, at least in an experiment in which a porous glass  filled with helium is in equilibrium with bulk superfluid.  
\\ \indent 
Naturally, there is an objective limit to how closely the simple, spherical model of confinement considered here can reproduce results of experiments aimed at investigating superfluid properties of helium inside porous media, whose microscopic structure is typically one of interconnected cylindrical channels. However, this result is consistent with experimental evidence indicating that a characteristic confining length of the order of 25 \AA\ may correspond to the lower limit of existence of superfluidity of \he4 in porous media.\cite{shira}
\\ \indent
Fig. \ref{rsofT} shows our computed values of the superfluid fraction $\rho_S(T)$ as a function of temperature, for a cluster comprising $N=45$ \he4 atoms, enclosed in the two spherical cavities considered here. Also shown for comparison is the corresponding value for a free \he4 cluster.\cite{unp} Although the superfluid signal is weaker than that which one would observe in a finite cluster, nonetheless it remains robust, comparable values of the superfluid fraction being observed only at a fraction of a K lower temperature. 
Within the statistical uncertainties of the calculation, the estimates obtained for this specific cluster are indistinguishable, This is peculiar, considering that the structure of the two clusters is very different, as shown in Fig. \ref{hig}. On a glass substrate, the character of the superfluid transition can be expected to be largely 2D; in the Li cavity, on the other hand, there are two shells, with a rather clear demarcation between the two, even though, as remarked above, some finite atomic overlap exists. \\ \indent
One might be inclined to think that in the less attractive Li cavity the superfluid response is essentially that of two almost independent 2D shells, or radii $\sim 3$ \AA\ and $\sim$ 6 \AA.  However, it should be noted that the numbers of particles in the two shells are 9 and 36 respectively, i.e., the effective 2D density is approximately 0.08 \AA$^{-2}$ for both, i.e., well into the crystalline region of the bulk phase diagram.\cite{gordillo} Moreover, the rapid decrease of the superfluid response for $N > 45$ concides with the sharp reduction of the overlap between the two shells, signaling that quantum-mechanical exchanges of atoms in different shells play a vital role in underlying the superfluid response of these clusters. We computed the local superfluid density\cite{fabio} and consistently found it to be homogeneous throughout the cluster, i.e., not concentrated mostly in some parts of it (e.g., one of the two shells).
This physical behaviour closely resembles that observed in simulations for parahydrogen clusters trapped inside nanoscale spherical cavities.\cite{tokunbo}
\subsection{Momentum Distribution}
 \begin{figure}  [h]        
\centerline{\includegraphics[scale=0.38]{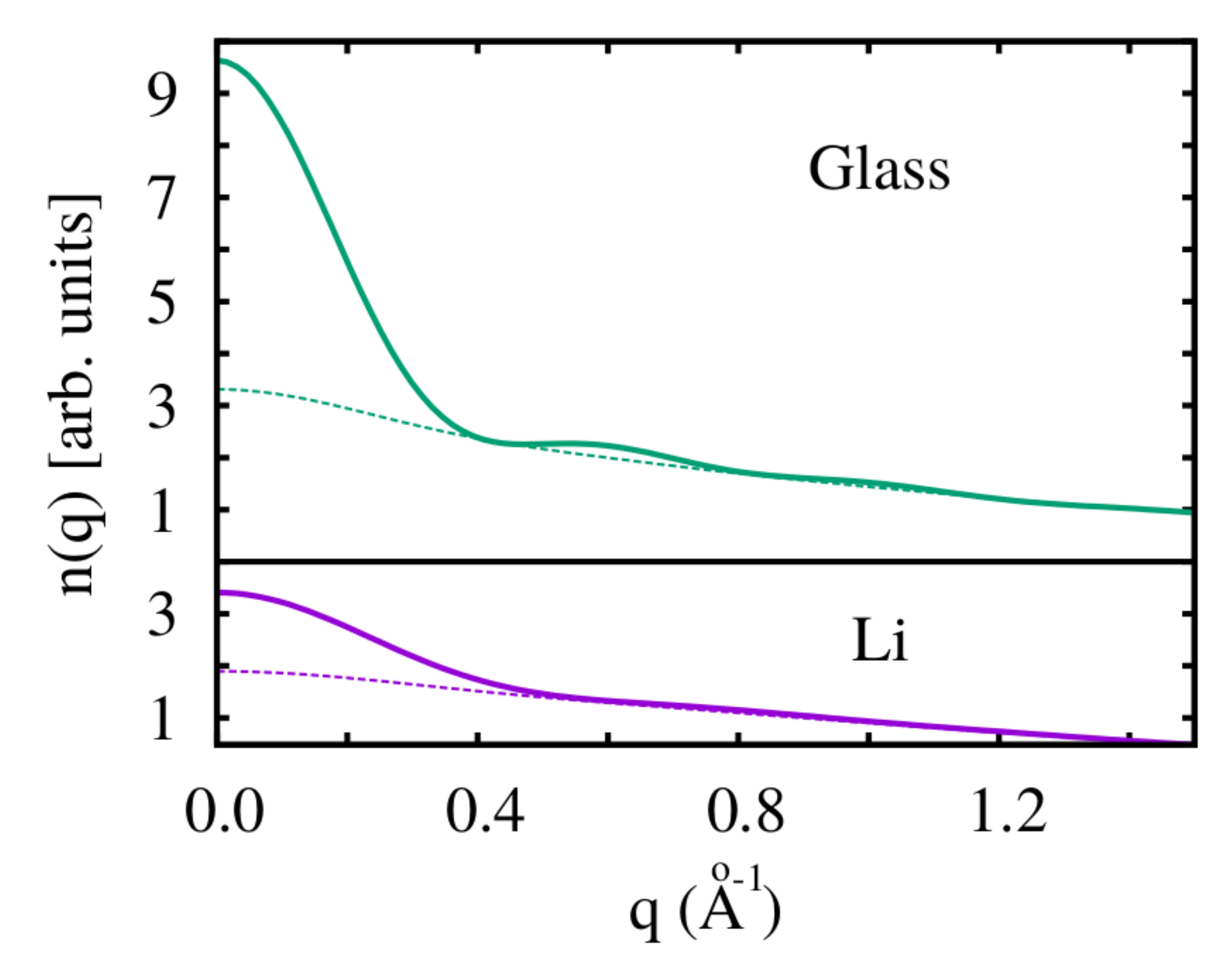}} 
\caption{Spherically averaged momentum distribution $n(q)$ computed for a cluster of $N=45$ \he4 atoms enclosed inside a  Li (lower panel) and glass (upper panel) cavity of radius $R=10$ \AA. Dashed lines refer to data at temperature $T$=2 K, solid lines at $T$=0.5 K.}
\label{nq}
\end{figure} 
As mentioned in the introduction, one of the main reasons for studying the physics of \he4 clusters in confinement is that one might be able to access their superfluid response more easily and directly than by performing spectroscopy of molecular probes embedded in free clusters. For example, one may measure the momentum distribution, typically by scattering neutrons off a sample of \he4 adsorbed inside some suitably chosen porous medium. On the assumption that the bulk of the signal should come from \he4 inside the cavities, one may 
 look for the appearance at low temperature of a peak at zero momentum,\cite{sokol} which signals the onset of Bose-Einsten condensation, intimately connected to superfluidity. 
\\ \indent
Fig. \ref{nq} shows the spherically average single-particle momentum distribution $n(q)$, computed for a cluster of 45 \he4 atoms inside the two cavities that we considered in this work, at the two temperatures $T$=2 K and $T$=0.5 K. The appearance of a peak at short momenta as the temperature is lowered  below $\sim 1$ K, i.e., when the clusters turn superfluid, is clear. The peak is not sharp as it would be in bulk superfluid, but rather broadened by the fact the system is confined over a length of $\sim$ 1 nm. In contrast, no such a peak develops if crystallization occurs inside the cavities. There is a quantitative difference between the signal observed in the glass cavity, inside which superfluidity is essentially 2D (as mentioned above), and in the Li one. The peak inside the glass cavity is stronger,  $n(q)$ slightly more structured and  falls off more slowly at high $q$ than inside the Li cavity, due to the localization of atoms in the vicinity of the cavity surface.
\section{Conclusions}
The purpose of this study has been that of gaining insight into the physical behaviour of nanoscale \he4 clusters inside a spherical cavity, with particular attention to the superfluid properties. The idea is that of providing a possible, alternate avenue to the exploration of clusters, one that does not involve the use of a foreign probe embedded in the cluster. Obviously, the presence of the confining surface has a significant effect on the structure of the clusters, but the main result of this work is that superfluidity can still be observed at temperatures comparable to those at which its onset is predicted for free clusters.
\\ \indent
We have focused on spherical cavities of radius 1 nm, because they can accomodate clusters of a few tens of atoms, which is an interesting size because of the strong competition between surface and bulk energetics. We have shown that in a cavity carved inside a medium like silica, which exerts a strong attraction on \he4 atoms, superfluidity can only occur in the form of a film coating the surface of the cavity, at equilibrium density. In an experiment in which the system is in thermal contact with a bath of superfluid \he4, our calculations show that cavities will be filled with frozen \he4, i.e., no superfluid signal will be seen, at least down to $T=0.5$ K. On the other hand, in a weakly attractive cavity, especially one like that studied here in which the repulsive  core of the interaction between atoms and cavity wall extends as far out  as a third of the radius of the cavity itself, then one can expect the \he4 fluid inside the cavity to display a significant superfluid response.
\\ \indent
Clearly, the mathematical model contains important simplifications, notably the assumption that the cavity may be regarded as perfectly spherical and smooth. The expected minimum distance from the surface (over 3 \AA) at which atoms sit in the case of a Li substrate seems to justify at least in part the neglect of corrugation and surface defects, an assumption routinely made in numerical studies of adsorption of He or parahydrogen on alkali substrates.\cite{io} In the case of a strongly attractive substrate such as glass, such an assumption is clearly far more questionable.
It is our hope that this work will provide at least some general aid in the design of experiments aimed at probing the superfluid response of nanoscale clusters of helium, or other quantum fluids (e.g., parahydrogen) in confinement.

\section*{Acknowledgments}
This work was supported in part by the Natural Science and Engineering Research Council of Canada. Computing support from Westgrid is gratefully acknowledged.


\begin{thebibliography}{99}
\bibitem{note}
It should be noted that fabrication techniques with the potential of extending significantly the diameter of the cages are being developed. See, for instance, W. Roth {\em et al.}, Nature Chemistry {\bf 5}, 628 (2013).
\bibitem{unp}
The calculation for the specific free \he4 clusters mentioned in this work was carried out in this work separately. Because such a small cluster is very loosely bound, an artificial confining potential was utilized, of the form $V(r)={\rm tan}(\pi r/2R)$, with $R$=20 \AA\ and $R$=25 \AA, yielding consistent results.
\end{thebibliography}
\end{document}